\documentclass[rotate,twocolumn,showpacs,preprintnumbers,amsmath,amssymb]{revtex4}

\usepackage{rotate}
\usepackage{epsfig}
\usepackage{graphicx}
\usepackage{dcolumn}
\usepackage{bm}

\def\DESepsf(#1 width #2){\epsfxsize=#2 \epsfbox{#1}}

\begin{document}


\title{\boldmath
{Dynamically enhanced $B\to K^*\pi$ decays in perturbative QCD}}

\author{Yong-Yeon Keum}
\affiliation{%
EKEN Lab. Department of Physics \\
Nagoya University, Nagoya 464-8602 Japan \\
{\it Email: yykeum@eken.phys.nagoya-u.ac.jp}
}%

\date{\today}

\begin{abstract}
We investigate the $B \to K^{*}\pi$ decays, one of hardly
understandable processes among charmless B-meson decays,
within the perturbative QCD method. Owing to the dynamically enhanced
mechanism in PQCD, we obtain large branching ratios and large direct
CP asymmetries:
$Br(B^0 \to K^{*\pm}\pi^{\mp})=(9.1^{+4.9+0.3}_{-3.9-0.2})\times 10^{-6}$,
$Br(B^{\pm} \to K^{*0}\pi^{\pm})=(10.0^{+5.3}_{-3.5}\pm 0.0)\times 10^{-6}$;
$Acp(B^0 \to K^{*\pm}\pi^{\mp})=(-19.2^{+0.5}_{-1.7})\%$,
and $Acp(B^{\pm} \to K^{*\pm}\pi^{0})=(-43.7^{+4.0}_{-4.2})\%$.
The branching ratios are consistent with experimental data and
large direct CP violation effects will be tested by near future
experimental measurements in Asymmetric B-factory.
\end{abstract}

\maketitle


The predictive power of the
perturbative QCD (pQCD) approach has been demonstrated 
successfully in exclusive 2-body B-meson decays, 
especially charmless
B-meson processes \cite{LiYu,KLS01,LUY,LuYa,phiK,phiKst,KK,keum}
which is based on $k_T$
factorization theorem \cite{BS}. 
This is a modified version of the pQCD theory for
exclusive processes\cite{BL}. 
The idea is to
separate hard scattering kernels from a high-energy QCD process, which are
calculable in a perturbative way . Nonperturbative parts are organized
into universal hadron distribution amplitudes, 
which can be determined from experimental data. 
By introducing parton transverse momenta $k_{\bot}$, 
we can generate naturally the Sudakov suppression effect 
due to the resummation of large double 
logarithms $Exp[-{\alpha_s C_F \over 4 \pi} \ln^2({Q^2\over k_{\bot}^2})]$,
which suppress the long-distance contributions in the small $k_{\bot}$ region
and give a sizable average $<k_{\bot}^2> \sim \bar{\Lambda} M_B$. 
This can resolve the end point singularity problem and 
allow the applicability of pQCD to exclusive B-meson decays. 
We found that almost all of the contributions to the matrix element
come from the integration region where $\alpha_s/\pi < 0.3$ and 
the pertubative treatment can be justified.

In the pQCD approach, we can predict the contribution of non-factorizable
term and annihilation diagram on the same basis as the factorizable one.
A folklore for annihilation contributions is that they are negligible
compared to W-emission diagrams due to helicity suppression. 
However the operators $O_{5,6}$ with helicity structure $(S-P)(S+P)$
are not suppressed and give dominant imaginary values, 
which is the main source of strong phase in the pQCD approach.

An alternative method to exclusive $B$ meson decays is QCD-factorization 
approach (QCDF)\cite{BBNS}, which is based on collinear factorization theorem.

For some modes, such as the $B\to K\pi$ decays, the difference between
the pQCD and QCDF approaches may not be signigicant. 
As explained in ref.\cite{KLS01}, 
the typical hard scattering scale is about 1.5 GeV.
Since the RG evolution of the Wilson coefficients $C_{4,6}(t)$ increase
drastically as $t < M_B/2$, while that of $C_{1,2}(t)$ remain almost
constant, we can get a large enhancement effects from both wilson
coefficents and matrix elements in pQCD. 
 
In general the amplitude can be expressed as
\begin{equation}
Amp \sim [a_{1,2} \,\, \pm \,\, a_4 \,\,
\pm \,\, m_0^{P,V}(\mu) a_6] \,\, \cdot \,\, <K\pi|O|B>
\label{eq:2}
\end{equation}
with the chiral factors $m_0^P(\mu)=m_P^2/[m_1(\mu)+m_2(\mu)]$ for
pseudoscalar meson 
and $m_0^{V}= m_V$ for vector meson.
To accommodate the $B\to K\pi$ data in the factorization and
QCD-factorization approaches, one relies on the chiral enhancement by
increasing the mass $m_0$ to as large values about 3 GeV at $\mu=m_b$ scale.
So two methods accomodate large branching ratios of $B \to K\pi$ and
it is difficult for us to distinguish two different methods in $B \to
PP$ decays. In addition,
the direct CP asymmetries in $B\to K\pi$ decays are not large enough to
distinguish the two approaches after taking into account the theoretical
uncertainties. 

However the difference can be detected in the direct CP asymmetry
of $B^0 \to \pi^{\pm}\pi^{\mp}$ process because of the different
power counting rules
and the branching ratios of
in $B \to PV$ modes since there is no
chiral enhanced factor in LCDAs of the vector meson.
Due to the different power counting rules of the QCDF 
and pQCD approaches, based on collinear and $k_T$ factorizations, 
respectively, the vertex correction
is the leading source of the strong phase in the former, and the
annihilation diagram is in the latter. The strong phases derived from the
above two sources are opposite in sign, and the latter has a large
magnitude. This is the reason QCDF prefers a small and positive CP
asymmetry $C_{\pi\pi}$ \cite{Ben}, while pQCD prefers a large and negative
$C_{\pi\pi}\sim -30\%$ \cite{KLS01,keum}.

We can test whether dynamical enhancement 
or chiral enhancement is responsible
for the large $B \to K\pi$ branching ratios 
by measuring the $B \to \phi K^{(*)}$ modes.
In these modes penguin contributions dominate, 
such that their branching ratios are
insensitive to the variation of the unitarity angle $\phi_3$.
According to recent
works within QCDF\cite{CY}, 
the branching ratio of $B \to \phi K$ is $(2-7)
\times 10^{-6}$ including $30\%$ annihilation contributions in real
part of amplitudes
within QCD-factorization approach. 
However pQCD predicts $10 \times 10^{-6}$ \cite{phiK} with
mostly pure imaginary annihilation contributions.   
For $B \to \phi K^{*}$ decays, QCDF gets about $9 \times 10^{-6}$\cite{CKY},
but pQCD have $15 \times 10^{-6}$\cite{phiKst}.
Because of these relatively small branching ratios for $B\to PV$ and $VV$ decays
in QCD-factorization approach, they can not globally fit the
experimental data for $B\to PP,VP$ and $VV$ modes simultaneously
with same sets of free parameters $(\rho_H,\phi_H)$ and $(\rho_A,\phi_A)$
\cite{zhu}. To expalin large branching ratios of $B \to VP$ modes,
they have to break the universality of free parameter sets with
$\rho_{H,A}\geq 1$ and finally lost the predictive power.

In this letter we investigate the more complicated $B\to K^*\pi$
processes, which contain both tree and penguin contributions, 
while $B \to \phi K^{(*)}$ is a pure penguin process. 
It is well known that it is very 
difficult to explain the observed $B\to K^*\pi$ branching ratios using the 
factorization assumption (FA) \cite{BSW} and QCDF\cite{zhu}: 
the experimantal measurements are much larger than the theoretical
predictions. 

The reason is as follows. 
The measured $Br(B\to K^*\pi)$ are roughly the same as 
$Br(B\to K\pi)\sim 2\times 10^{-5}$. However, the penguin operators
$O_{5,6}$ contribute to the latter, but not to the former. Due to the
loss of this important piece of contributions, the predicted
$Br(B\to K^*\pi)$ become a quarter of the predicted $Br(B\to K\pi)$. The
same difficulty has been encountered in the $B\to\rho\pi$ decays, where
the vector meson is replaced by a $\rho$ meson \cite{LuYa}. This
controversy remains, no matter how the angle $\phi_3$ is varied \cite{WS}.
Hence, the $B\to K^*\pi$ decays is worth of an intensive study.

We shall evaluate the branching ratios of the following modes,
\begin{eqnarray}
& &B^\pm\to K^{*0}\pi^\pm\;,\;\;\;\;B_d^0\to K^{*\pm}\pi^\mp\;, 
\nonumber\\
& &B^\pm\to K^{*\pm}\pi^0\;,\;\;\;\;B_d^0\to K^{*0}\pi^0\;,
\label{bkpi}
\end{eqnarray}
and the CP asymmetries, for instance, $f=K^{*+}\pi^{-}$
\begin{equation}
A_{CP}=\frac{{\rm Br}({\bar B}_d^0\to \bar{f})
- {\rm Br}(B_d^0\to f)}
{{\rm Br}({\bar B}_d^0\to \bar{f}) + 
{\rm Br}(B_d^0\to f)}\;,
\label{cp1}\\
\end{equation}
as functions of the unitarity angle $\phi_3$. It will be shown that
penguin and annihilation amplitudes in the $B\to K^*\pi$ decays are
greatly enhanced by Wilson evolutin effects. There is also small
enhancement from the $K^*$ meson wave functions, which are more asymmetric
than the kaon wave funcitons, and from the $K^*$ meson decay constant
$f_{K^*}$, which is larger than the kaon decay constant $f_K$. It turns
out that these enhancements compensate the loss of the $O_{5,6}$
contributions, and that PQCD predictions are in agreement with the data.

The decay rates of $B\to K^{*}\pi$ have the expressions,
\begin{equation}
\Gamma=\frac{G_F^2M_B^3}{128\pi}|{\cal A}|^2\;.
\label{dr1}
\end{equation}
The decay amplitudes ${\cal A}$ for the different modes are written as
\begin{eqnarray}
{\cal A}({K^{*0}\pi^+})&=&f_{K^*}V_t^*F^{P(d)}_{e}
+V_t^*{\cal M}^{P(d)}_e
+f_BV_t^*F^{P(u)}_a      \nonumber \\
&& +V_t^*{\cal M}^{P(u)}_a
-f_BV_u^*F_a^T-V_u^*{\cal M}_a^T\;,
\label{Map}\\
{\cal A}({K^{*+}\pi^-})&=&f_{K^*}V_t^*F^{P(u)}_{e}
+V_t^*{\cal M}^{P(u)}_e
+f_BV_t^*F^{P(d)}_a \nonumber \\
&& +V_t^*{\cal M}^{P(d)}_a
-f_{K^*}V_u^*F_e^T-V_u^*{\cal M}_e^T\;,
\label{Mbp}\\
\sqrt{2}{\cal A}({K^{*+}\pi^0})&=&f_{K^*}V_t^*F^{P(u)}_{e}
+V_t^*{\cal M}^{P(u)}_e 
+f_BV_t^*F^{P(u)}_a \nonumber \\
&& +V_t^*{\cal M}^{P(u)}_a
+f_\pi V_t^*F^P_{eK}+V_t^*{\cal M}^P_{eK}
\nonumber\\
& &-f_{K^*}V_u^*F_{e}^T-V_u^*{\cal M}_e^T
-f_BV_u^*F_a^T \nonumber \\
&& -V_u^*{\cal M}_a^T
-f_\pi V_u^*F_{eK}^T-V_u^*{\cal M}_{eK}^T \;,
\label{Mapp}\\
\sqrt{2}{\cal A}({K^{*0}\pi^0})
&=&f_{K^*}V_t^*F^{P(d)}_{e}
+V_t^*{\cal M}^{P(d)}_e
+f_BV_t^*F^{P(d)}_a \nonumber \\
&& +V_t^*{\cal M}^{P(d)}_a
+f_\pi V_t^*F^P_{eK} + V_t^*{\cal M}^P_{eK}
\nonumber\\
& &-f_\pi V_u^*F_{eK}^T-V_u^*{\cal M}_{eK}^T \;,
\label{Mbpp}
\end{eqnarray}
to which ${\cal A}({K^{*0}\pi^-})$, ${\cal A}({K^{*-}\pi^+})$,
$\sqrt{2}{\cal A}({K^{*-}\pi^0})$, and
$\sqrt{2}{\cal A}({{\bar K}^{*0}\pi^0})$ are identical, respectively, but
with the the product $V_t^*$ ($V_u^*$) of the CKM matrix elements replaced
by $V_t$ ($V_u$).
\begin{table}[b!]
\caption{Contribution to the $B_u^{+} \to K^{*+} \pi^{0}$ decay from
each form factor and nonfactorizable amplitude. 
 \label{table-1} }
\begin{ruledtabular}
\begin{tabular}{|c||c|}
~~~~~~~~$f_{K^{*}}\,F_e^T$~~~~~~~~ &  $-1.202 \times 10^{-1}$    \\
~~~~~~~~$f_{K^{*}}\,F_e^P$~~~~~~~~ & $4.424 \times 10^{-3}$   \\
$f_{B}\,F_a^T$ & $-3.521\times 10^{-2} + \,\, i \,\, 4.422\times 10^{-3}$~~~~~ \\
$f_{B}\,F_a^P$ & $2.002\times 10^{-3} - \,\, i \,\,  2.076\times 10^{-3}$~~~~~\\ 
\hline
$M_e^T$ & $-4.369\times 10^{-3} + \,\, i \,\, 5.317\times 10^{-3}$~~~~~  \\      
$M_e^P$ & $1.733\times 10^{-4} - \,\, i \,\,  2.612\times 
10^{-4}$~~~~   \\       
$M_a^T$ & $-4.413\times 10^{-4} + \,\, i \,\, 1.297\times 10^{-3}$~~~~~   \\ 
$M_a^P$ & $-3.355\times 10^{-5} - \,\, i \,\, 4.591\times
10^{-6}$~~~~~\\ 
\hline      
$f_{\pi}\,F_{eK}^T$ &  $-1.089 \times 10^{-2}$    \\
$f_{\pi}\,F_{eK}^P$ & $-1.832 \times 10^{-3}$   \\
$M_{eK}^T$ & $-1.432\times 10^{-2} - \,\, i \,\, 3.226\times
10^{-3}$~~~~~ \\      
$M_{eK}^P$ & $-8.466\times 10^{-5} - \,\, i \,\, 1.894\times 10^{-5}$~~~~~  \\
\end{tabular}
\end{ruledtabular}
\end{table}
\begin{table*}[t!]
\caption{Enhancement effects in the  $B^{\pm} \to K^{*0} \pi^{\pm}$
decay amplitudes  
\label{table-2} }
\begin{ruledtabular}
\begin{tabular}{|c||c|c||c|c||} 
\multicolumn{1}{|c||}{~~~Scales~~~} & 
\multicolumn{2}{c||}{$\mu=t$~~~($\sim 1.5$ GeV)~~~~~~~} & 
\multicolumn{2}{c||}{$\mu = m_b/2$~~~~~~~}  \\ \hline \hline
Amplitudes  & Re~($10^{-4}$ GeV)~~~ & Im~($10^{-4}$ GeV)~~ 
& Re~($10^{-4}$ GeV)~~ &  Im~($10^{-4}$ GeV)~~\\ \hline \hline
$f_{K^*} F_{e}^T$ & -1202.0  &  ---  & -1163.0  & ---  \\ \hline
$f_{K^*} F_{e}^{P}$ &  45.9  &  ---  & 35.6  & ---  \\ \hline
$f_{B}\,\, F_{a}^T$ & -350.1  & 44.2  & -340.7  & 42.8  \\ \hline
$f_{B}\,\, F_{a}^{P}$ & 20.0  & -20.7  & 15.5 & -14.8  \\ \hline
   $ M_{e}^T$  & -43.7  & 53.2    & -33.8  & 41.1  \\ \hline
   $ M_{e}^{P}$  & 3.1  & -4.3    & 2.4  & -3.3  \\ \hline
   $ M_{a}^T$  & -4.4  & -13.0  & -3.4  & 10.2  \\ \hline
   $ M_{a}^{P}$  & -0.3  & 0.0   & -0.3  & 0.0  \\ \hline \hline
\multicolumn{1}{|c||}{Br. with ann.} 
& \multicolumn{2}{c||}{ $ 10.0 \times 10^{-6} $ } 
& \multicolumn{2}{c||}{ $  6.0     \times 10^{-6} $ } \\ \hline
\multicolumn{1}{|c||}{Br. without ann.} 
& \multicolumn{2}{c||}{ $ 4.6 \times 10^{-6} $ } 
& \multicolumn{2}{c||}{ $  2.8     \times 10^{-6} $ } \\ 
\end{tabular}
\end{ruledtabular}
\end{table*}

The detail expression of analytic formulas for all amplitudes
($F_{i}^{T,P}$ and $M_{i}^{T,P}$) will be presented elsewhere\cite{KL}.
In the above expressions $f_B$ is the $B$ meson decay constant. The
notations $F$ represent factorizable contributions (form factors), and
${\cal M}$ represent nonfactorizable (color-suppressed) contributions. The
subscripts $a$ and $e$ denote the annihilation and W-emission topology,
respectively. The superscript $P(T)$ denotes contributions from the penguin
(Tree) operators. $F_{a(e)}^T$, associated with the time-like $K^*$-$\pi$ form
factor ($B\to\pi$ form factor), and ${\cal M}_{a(e)}^T$ are from the
operators $O^{(u)}_{1,2}$. The factorizable contribution $F^P_{eK}$
($F_{eK}^T$) is associated with the $B\to K^{*}$ form factor from the
penguin (tree) operators, and ${\cal M}^P_{eK}$ (${\cal M}_{eK}^T$) is the
corresponding nonfactorizable contribution. 

In our numerical analysis, 
we use $G_F=1.16639\times 10^{-5}$ GeV$^{-2}$, the Wolfenstein
parameters $\lambda=0.2196$, $A=0.819$, and $R_b=0.38^{+0.10}_{-0.06}$ 
for the CKM matrix elements, the masses $M_B=5.28$ GeV
and $M_{K^*}=0.892$ GeV, and ${\bar B}_d^0$ ($B^-$) meson lifetime
$\tau_{B^0}=1.55$ ps ($\tau_{B^-}=1.65$ ps) \cite{LEP}. The angle
$\phi_3=80^o$ was extracted from the data of the $B\to K\pi$ and $\pi\pi$
decays \cite{KLS01,keum}. With all the meson wave functions given in our
previous works, we calculate the contributions from all the topologies
as shown in Figs. 2 and 3 in $B \to K\pi$ paper\cite{KLS01}.
The allowed range of B-meson shape parameter, 
$0.36\,\, GeV < \omega_B < 0.44 \,\, GeV$,
and chiral factor for pion, $1.2 \,\, GeV < m_{0}^{\pi} < 1.6 \,\,
GeV$ is determined from the reasonable $B\to \pi$ and $B \to K$ 
transition form factors. 
\begin{table}[t!]
\caption{PQCD predictions and experimental data for the
$B \to K^{*}\pi$ branching Ratios in unit of $10^{-6}$. 
\label{table-3} }
\begin{ruledtabular}
\begin{tabular}{|c||ccc||c|}
Modes & CELO & Belle & BaBar~& PQCD \\ \hline
$K^{*\pm}\pi^{\mp}$ & $16^{+6.3}_{-5.4}$  &
$26.0\pm 9.0$ & --- & $9.1^{+4.9+0.3}_{-3.9-0.2}$ \\ \hline
$K^{*0}\pi^{\pm}$   & $< 16$ &
$16.2^{+4.8}_{-4.5}$ &
$15.5\pm 3.8$ & $10.0^{+5.3}_{-3.5}\pm 0.0$ \\ \hline
$K^{*\pm}\pi^{0}$ & --- & --- & --- & $3.2^{+1.9+0.6}_{-1.2-0.2}$ \\ \hline
$K^{*0}\pi^{0}$   & --- & --- & --- & $2.8^{+1.6}_{-1.0}\pm0.0$ \\
\end{tabular}
\end{ruledtabular}
\end{table}
\begin{figure}[b!]
\includegraphics[angle=-90,width=0.3\textwidth]{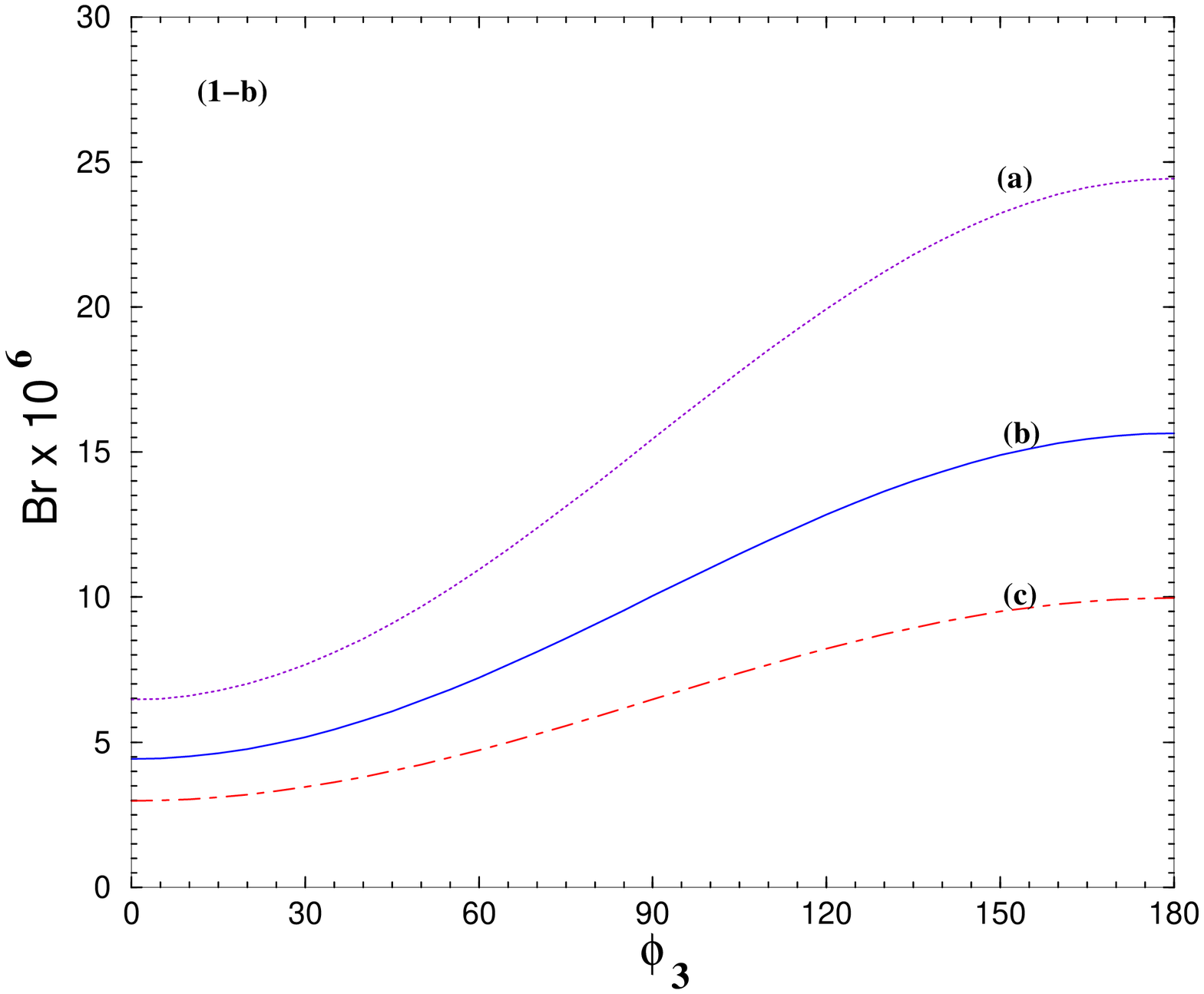} 
 \includegraphics[angle=-90,width=0.3\textwidth]{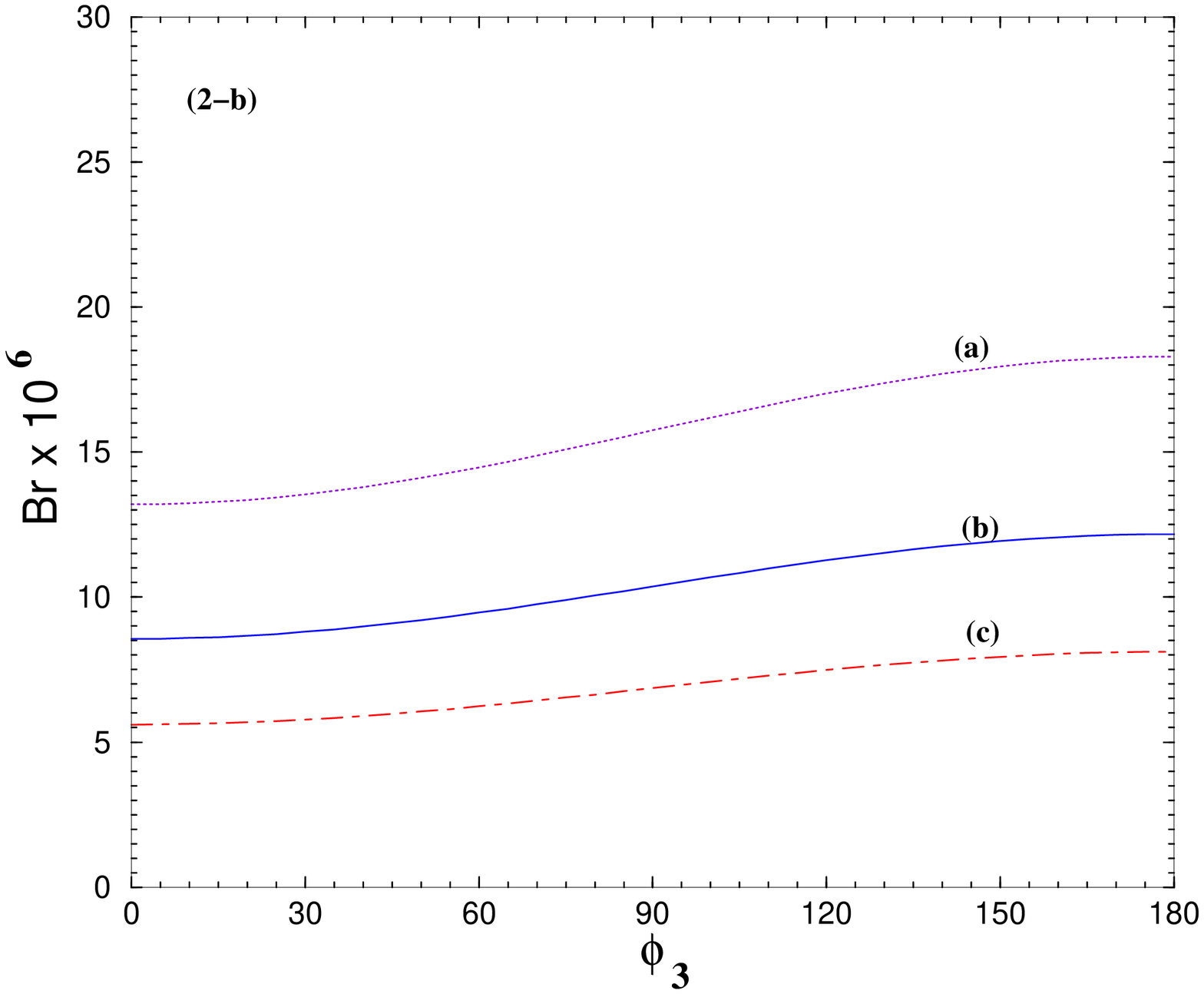} 
\caption{Branching ratios of $B^0\to K^{*\pm}\pi^{\mp}$(1-b)
and $B^0\to K^{*0}\pi^{\pm}$ (2-b) versus $\phi_3$ angle.
The dot-, solid- and dashed-dot line stands for the
averaged branching ratio with 
(a) $\omega_B=0.36 \,\,GeV$ and $m_0^{\pi}=0.16 \,\, GeV$, 
(b) $\omega_B=0.40 \,\,GeV$ and $m_0^{\pi}=0.14 \,\, GeV$, and
(c) $\omega_B=0.44 \,\,GeV$ and $m_0^{\pi}=0.12 \,\, GeV$,
respectively.
\label{fig-1}}
\end{figure}
For example, all amplitudes for the $B^{+} \to K^{*+} \pi^{0}$
modes are listed in Table \ref{table-1}, whose values are mostly 
the same magitude for other decay channels, because the difference
comes only from electroweak penguin contributions. 
We show in Table \ref{table-2} the
enhancing effect by comparing the decay amplitudes evaluated at the
characteristic hard scales $t$ in PQCD and $m_b/2$ in QCDF. It is also
found that the annihilation contributions are sizable in
two-body charmless $B$ meson decays for the heavy-meson mass around 5 GeV
\cite{KLS01} and in fact contributed about 60\% fraction of the
branching ratios, since factorized annihilation penguin
contribution has large imaginary part and also the same order of
magnitudes in real part as one of the factorized penguin contribution.
As expected, the dominant factorizable penguin amplitudes are
enhanced by about 30\% due to the Wilson evolution, more than the
factorizable tree amplitudes are.
\begin{figure}[t!]
\includegraphics[angle=-90,width=0.3\textwidth]{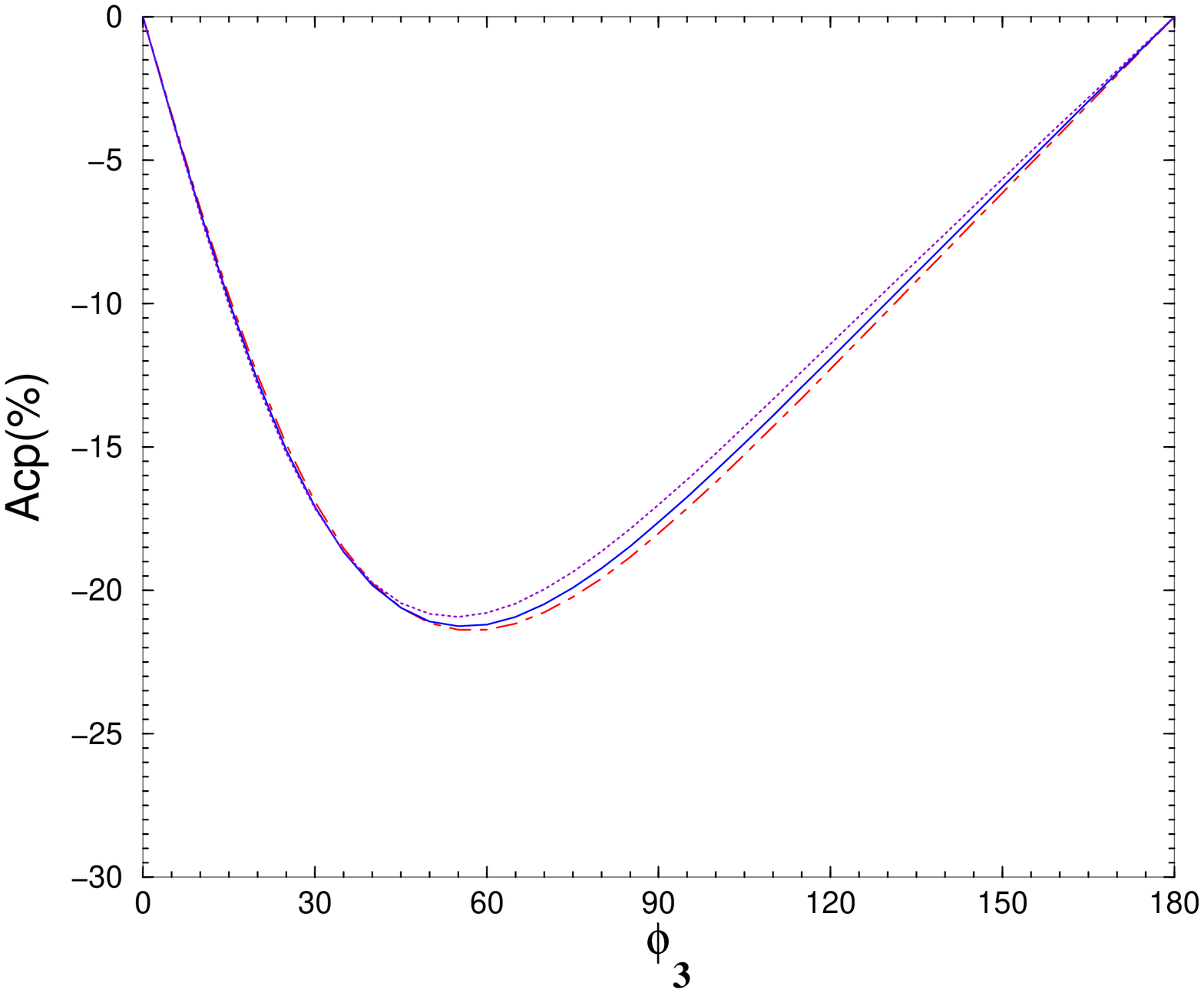} 
 \includegraphics[angle=-90,width=0.3\textwidth]{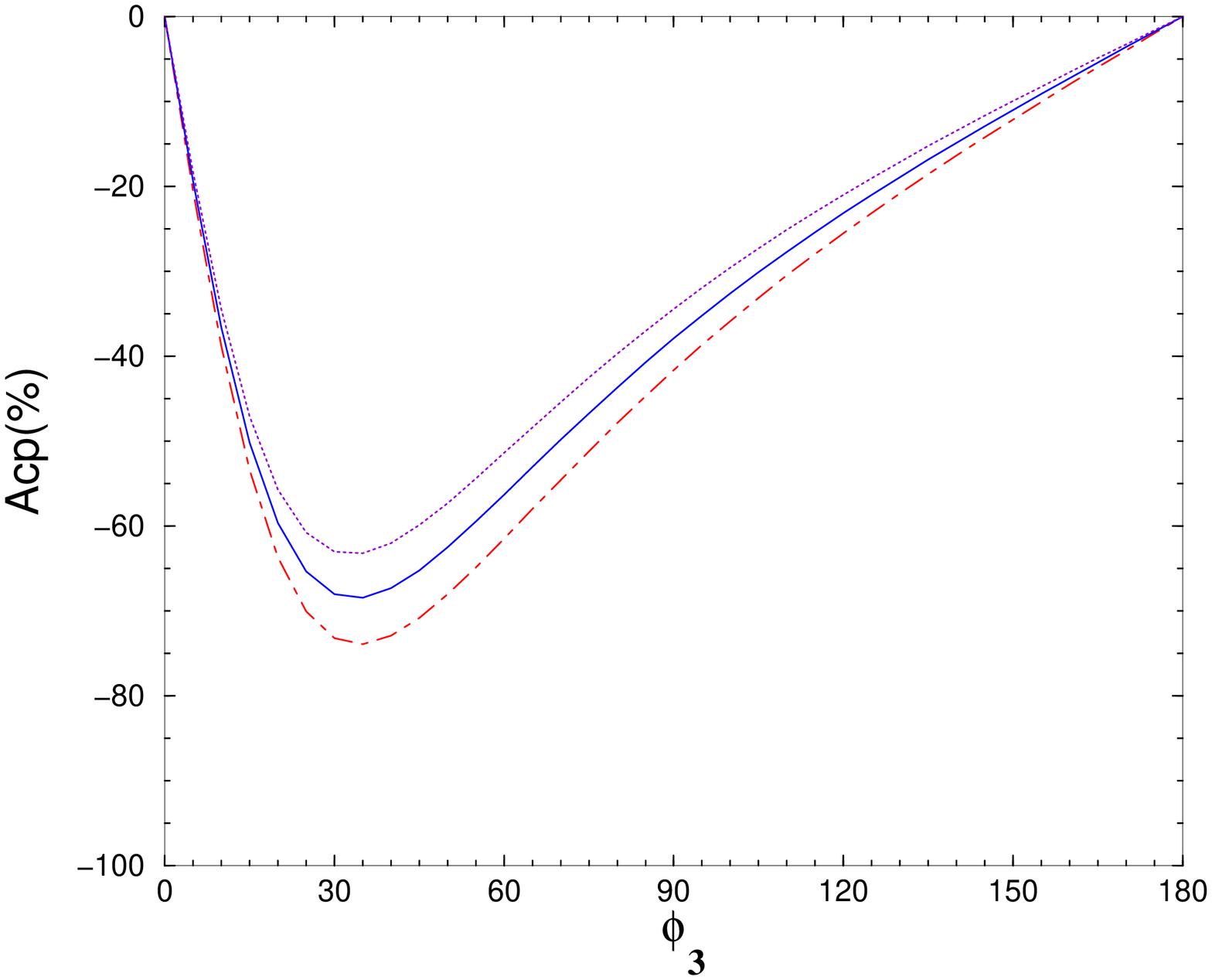} 
\caption{Direct CP asymmetry in $B^0 \to K^{*\pm}\pi^{\mp}$(upper)
and $B^{\pm}\to K^{*\pm}\pi^{0}$(lower).
\label{fig-2}}
\end{figure}
The PQCD predictions for the $B\to K^*\pi$ branching
ratios, presented in Table \ref{table-3}, are consistent with 
present experimental data. Here the first uncertainty comes from the
allowed ranges of both $\omega_B$ and $m_0^{\pi}$, and
the second one comes from the uncertainty of 
$R_b=|V_{ub}/V_{cb}|/\lambda$.

Our predictions for the CP asymmetries in
$B\to K^*\pi$ decays are given in Table \ref{table-4}, 
which have the same sign
as of those in $B\to K\pi$ decays. For $\phi_3=80^o$,
CP asymmetries of $B^0\to K^{*\pm}\pi^{\mp}$ and
$B^{\pm}\to K^{*0}\pi^{\pm}$ become large due to the important imaginary
penguin annihilation amplitudes.

At last, the dependence of the branching ratios 
of $B^0\to K^{*\pm}\pi^{\mp}$ and $B^{\pm}\to K^{*0}\pi^{\pm}$
on the angle $\phi_3$ is shown in Fig.1
and large direct CP asymmetries of the
$B^0\to K^{*\pm}\pi^{\mp}$ and $B^{\pm}\to K^{*\pm}\pi^{0}$
on the angle $\phi_3$ is exhibited in Fig.2. 
The branching ratios of $K^{*\pm}\pi^{\mp}$ and $K^{*\pm}\pi^{0}$
increase with $\phi_3$ rapidly, 
while the $K^{*0}\pi^{\pm}$ and $K^{*0}\pi^{0}$  modes are
insensitive to the variation of $\phi_3$. The increase with $\phi_3$ is
mainly a consequence of the inteference between the penguin contribution
$F_e^P$ and the tree contribution $F_e^T$. The dependence on both the shape
parameter $w_B$ for the $B$ meson wave function 
and the chiral factor $m_0^{\pi}$ is also shown, which
is strong in the $K^{*\mp}\pi^{\pm}$ and $K^{*\mp}\pi^0$ modes, and weak
in the other two. The sensitivity is attributed to the fact that the
former contain both $F_e^P$ and $F_e^T$, which involve the $B$ meson
wave function, while the latter contain only $F_e^P$.
\begin{table}[ht!]
\caption{Direct CP Asymmetry of the $B \to K^{*}\pi$ decays.
Case A corresponds to $\omega_B=0.44 \,\, GeV$, $m_{0}^{\pi}=1.2 \,\,
GeV$, Case B is to $\omega_B=0.40 \,\, GeV$, $m_{0}^{\pi}=1.4 \,\,
GeV$, and  Case C is to $\omega_B=0.36 \,\, GeV$, $m_{0}^{\pi}=1.6 \,\,
GeV$. 
\label{table-4} }
\begin{ruledtabular}
\begin{tabular}{|c||c|c|c||c|}  
Modes & case A & case B &  case C & PQCD \\ \hline
$K^{*\pm}\pi^{\mp}$ & -19.6 \% & -19.2 \% & -18.7 \% &
$-19.2^{+0.5}_{-1.7}$ \% \\ \hline
$K^{*0}\pi^{\pm}$   & -4.4 \% & -3.6 \% & -2.7 \% & $
-3.6^{+0.9}_{-0.8}$ \% \\ \hline
$K^{*\pm}\pi^{0}$ & -47.9 \% & -43.7 \% & -39.7 \% &
$-43.7^{+4.0}_{-4.2}$ \% \\ \hline
$K^{*0}\pi^{0}$   & -10.7 \% & -9.6 \% & -8.8 \% &
$-9.6^{+0.8}_{-1.1}$ \% \\ 
\end{tabular}
\end{ruledtabular}
\end{table}

In this letter  in order to explain one of the hardly understandable
processes in charmless B-decays,
we have investigated the dynamical enhancement effect
in the $B\to K^*\pi$ decays within pQCD method. 
Owing to the dynamical enhancement of penguin contributions at $t\sim
1.5$ GeV, pQCD predictions for all the
$B\to K^*\pi$ modes are consistent with the present experimental data,
which is a crucial decay process to distinguish pQCD from other approachs.
We also predicted large direct CP asymmetry for $B^0 \to
K^{*\pm}\pi^{\mp}$ about -20\% and for $B^{\pm} \to
K^{*0}\pi^{\pm}$ about -40\%, which can be tested in near future
measurements. 
More detail works will appear elsewhere\cite{KL}.  


We wishes to thank S.J. Brodsky, H.Y. Cheng, H.-n. Li, A.I. Sanda 
and G. Zhu for helpful discussions.
This work was supported 
by Grant-in-Aid for Scientific Research from Ministry of Education,
Science and Culture of Japan.


\end{document}